\documentclass[letter,twocolumn,showpacs,preprintnumbers,amsmath,amssymb]{revtex4}
\usepackage{graphicx}% Include figure files
\usepackage{dcolumn}% Align table columns on decimal point
\usepackage{bm}% bold math

\begin{document}

%\preprint{APS/123-QED}

\title{Emergent Criticality from Co-evolution in Random Boolean Networks}
% Force line breaks with \\
\author{Min Liu}
\email{mliu3@uh.edu}
 \author{Kevin E. Bassler}%
 \email{bassler@uh.edu}
\affiliation{%
Department of Physics, University of Houston, Houston, TX 77204-5005
}%

\date{\today}% It is always \today, today,

\begin{abstract}
The co-evolution of network topology and dynamics is studied in an
evolutionary Boolean network model that is a
simple model of gene regulatory network. We find that a critical
state emerges spontaneously resulting from interplay between
topology and dynamics during the evolution. The final evolved state
is shown to be independent of initial conditions. The network
appears to be driven to a random Boolean network with uniform
in-degree of two in the large network limit. However, for
biologically realized network sizes, significant finite-size effects
are observed including a broad in-degree distribution and an average
in-degree connection between two and three. These results may be
important for explaining properties of gene regulatory networks.
\end{abstract}

\maketitle

\section{Introduction}
Boolean networks\cite{Kauffman1,Kauffman2,RBN1,RBN2,RBN5,RBN4,RBN3,SOCRBN}
have been extensively studied over the past three decades. They have
applications as models of gene regulatory networks, and also as
models of social and economic systems. 
As ``coarse-grained'' models of genetic networks
they aim to capture much of the observed systematic behavior of the networks while 
simplifying local gene expression to a binary (on/off) state\cite{lessmore}.
Despite their simplicity, recent work has demonstrated that the model
can indeed predict the essential features of the dynamics of a biological
genetic circuit\cite{lessE1,lessE2}. An important feature of
Boolean networks is that they have a continuous
phase transition between so-called ordered and chaotic phases. It has been
argued\cite{edge2,Kauffman3} that gene regulatory networks of living
systems should be at or close to criticality, at the so-called
``edge of chaos" between the two phases, because then they can maintain both the
evolvability and stability. 

Many studies of critical random Boolean networks(RBNs) have considered networks
with homogeneous topology
in which each node has the same number of inputs from other
nodes~\cite{Parisi,drossel2,drossel3,drossel4,RBN6,Socolar,economics,canalization2,canalization}.
Accumulating experimental
evidence\cite{experiment0,experiment1,experiment3}, however, shows
that real genetic networks do not have homogeneous connectivity,
but, instead, are topologically heterogeneous. This diversity of
architecture is, presumably, of great importance for the stability
of living cells. Studies of RBNs with heterogeneous topology have
analytically determined the location of the ordered to chaotic phase
transition in the large network limit and also
demonstrated how different kinds of topology affects the
stability of the dynamics\cite{analytic,Fox,Aldana1,Aldana2}. 
These studies emphasized
the importance of criticality and the influence of the network's
topology on its dynamics, but they did not attempt to explain
how critical networks with heterogeneous topology come to exist.

Recent analysis of real gene regulatory
networks\cite{experiment2} has uncovered the possibility that
the interactions between genes can change
in response to diverse stimuli.
The resulting changes in the network topology
can be far greater than what is expected simply from random mutation.
However, few general
principles are known about the evolution of network
topology. In an effort to determine what some of those principles may be
Bornholdt and Rohlf\cite{Bornholdt}
studied a simple model of neural networks known as
random threshold networks (RTNs). 
In their study, the topology of the RTN evolved according to a rule that
depends on the local dynamics of the network.
According to their rule, active nodes, whose binary state changes in time, tend to lose
links, while inactive node, whose binary state is fixed, tend to gain new links.
They observed that with this rule for changing network topology,
in the limit of large networks, 
the RTNs evolved to a critical network with average connectivity
$\overline{K}=2$. The study, therefore, discovered an
interdependence between
a network's dynamics and its topology.  

Motivated by 
these recent findings concerning gene regulatory networks and the behavior of
RTNs, here we
investigate 
the effect of a similar evolutionary rule on a simple
model of a genetic regulatory system.
In particular, 
we study an
evolutionary RBN model.
However, there is a fundamental difference between RTNs and RBNs.
In RTNs the dynamics of each node is controlled by the same threshold function, while in
RBNs the dynamics of each node is controlled by a randomly chosen Boolean function.
Therefore, in addition to evolving the topology of the network,
we also allow the Boolean functions used by the nodes to change. 
In the context of a model of gene regulation the rule we use to evolve the network topology 
assumes that
there is some selection pressure on an individual gene due to its activity. The
evolution causes genes that are in a frozen state of regulation, and, thus, are almost
nonfunctional, to gain functionality, while it reduces the functionality of genes that
are actively regulated. 
Thus this study investigates the co-evolution of network structure and network
dynamics.
We show that, independent of the initial topology of the
network, the RBN evolves to a critical network with a finite number of
nodes and which has a heterogeneous topology. Two different variants
of our model are considered, and our principal conclusions are the
same for both of them. Perhaps our most important result concerns
the finite-size effects of the model. As the size of the network
increases, the distribution of in-degree connectivity becomes
increasingly narrow and sharply peaked at a value of $K=2$. However,
for biologically realized network sizes, we show that the final
evolved critical state has a broad distribution of in-degree
connectivity with an average value between 2 and 3. Both of these
features also occur in real networks, suggesting that many of the
topological features of real networks may be due to their finite
size.

%%%%%%%%%%%%%%%%%%%%%%%%%%%%%%%%%%%%%%%%%%%%%

%%%%%%%%%%%%%%%%%%%%%%%%%%%%%%%%%%%%%%%%%%%%%
\begin{figure}[t]
\includegraphics[scale=0.52]{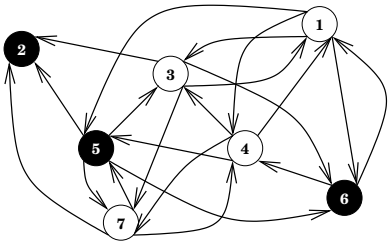}% Here is how to import EPS art
\caption{\label{rbn} A directed graph $G(7,3)$ represents a
homogeneous RBN with 7 nodes and in-degree connectivity of 3. Black
and white represent binary states `1' and `0' respectively. The
state vector of network is $\mathbf{\Sigma}(t) = (0,1,0,0,1,1,0)$.
The arrow on each link indicates the direction of information flow
in the network. }
\end{figure}
\section{Definition of Model}

\subsection{Dynamics of Random Boolean Networks }

A generalized RBN consists of $N$ randomly interconnected nodes,
$i=1, \cdots, N$, each of which has $K_i$ in-degree connections from
nodes that regulate its behavior. The simplest Boolean network model
is a homogeneous RBN in which each node has a same number of input
nodes. In this case, the connections between nodes is described by a
random directed graph $G(N,K)$ consisting of $N$ nodes with uniform
in-degree $K$. Figure~\ref{rbn} illustrates an example of $G(7,3)$.
Each node has a Boolean dynamical state at time $t$,  $\sigma_i(t) =
0 $ or $1$. The state of each node at time $t+1$ is a function of
all states of its $K_i$ regulatory nodes at time $t$. Hence, the
discrete dynamics of the network is given by
\begin{equation}\label{dynamics}
\sigma_i(t+1) = f_i\left(\sigma_{i_1}(t),\sigma_{i_2}(t),\cdots
,\sigma_{i_{K_i}}(t)\right)
\end{equation}
where $i_1$,$i_2$,$\cdots$,$i_{K_i}$ are those input nodes
regulating node $i$. The function $f_i$ is a Boolean function of
$K_i$ variables that determines the output of node $i$ for all of
the $2^{K_i}$ possible sets of input. 
Note that random Boolean functions can be generated with an ``interaction bias'' $p$
by setting the output value of the function for each set of input to be $1$ with probability
$p$.  
The bias $p$ can be interpreted as a biochemical reaction parameter.

%%%%%%%%%%%%%%%%%%%%%%%%%%%%%%%%%%%%%%%%%%

% Talk about the attractor and state space next.

%%%%%%%%%%%%%%%%%%%%%%%%%%%%%%%%%%%%%%%%%%

Given the Boolean state of each node $i$ at time $t$, $\sigma_i(t)$,
the state vector of network is defined as 
$\mathbf{\Sigma}(t)=(\sigma_1(t),\cdots,\sigma_N(t))$. 
The path
that $\mathbf{\Sigma}(t)$ takes over time $t$ is a dynamical
trajectory in the phase space of system.  
Because the
dynamics defined in Eq.~\ref{dynamics} is deterministic and the
phase space is finite, all dynamical trajectories eventually
become periodic. That is, after some possible transient behavior,
each trajectory will repeat itself forming a cycle given by,
\begin{equation}\label{attractor}
\mathbf{\Sigma}(t) = \mathbf{\Sigma}(t + \Gamma).
\end{equation}
The periodic part of the trajectory is the attractor of the
dynamics, and the minimum $\Gamma > 0$ that satisfies
Eq.~\ref{attractor} is the period of the attractor.

%In general, multiple attractors of different periods coexist in the
%phase space. The initial state $\mathbf{\Sigma}(0)$ determines which
%attractor the network finally reaches. The set of all initial states
%reaching the same attractor forms a basin in the phase space. 
Two phases exist in RBNs, chaotic and ordered, characterized by their
dynamical behavior\cite{RBN1,RBN2}. One important way of
distinguishing the two phases is to measure the
distribution of its attractor periods beginning with random initial
states. For RBNs in the chaotic phase the distribution of attractor
periods is sharply peaked near an average value that grows
exponentially with system size $N$, and for RBNs in the ordered
phase the distribution of attractor periods is sharply peaked near
an average value that is nearly independent of $N$. Critical RBNs,
however, have a broad power law distribution of attractor
periods\cite{Parisi}.

\subsection{Co-evolution in Random Boolean Networks}
In our model, the evolutionary changes of the topology of the
network are driven by the dynamics of the network, and the functions
that control the dynamics of network simultaneously evolve due to
the changes in the network topology. Thus, this study investigates
the co-evolution of network topology and dynamics. Similar to the
one used in Ref.~\cite{Bornholdt}, the topology-evolving rule is
simply that a frozen gene grows a link while an active gene loses a
link. The dynamical functions can be changed in either an annealed
or a quenched way. The detailed algorithm is defined as follows:
\smallskip
\begin{enumerate}
    \item Start with a homogeneous RBN, $G(N,K_0)$ with uniform in-degree
    connectivity $K_i=K_0$ for all $N$, and generate a random Boolean function $f_i$ for each node $i$.
    \item Choose a random initial system state $\mathbf{\Sigma}(0)$.
    Update the state using Eq.~\ref{dynamics}
    and find the dynamical attractor. See the appendix for a description
of the algorithm used to find the attractor.
    \item Choose a node $i$ at random and determine its average
    activity $\overline{O}(i)$ over the attractor.
    \begin{equation}\label{criterion}
    \overline{O}(i) = \frac{1}{\Gamma} \sum_{t=T}^{T+\Gamma-1} \sigma_i(t)
\end{equation}
where $T$ is a time large enough so that the periodic attractor has
been reached, and $\Gamma$ is the period of the attractor. If
$\overline{O}(i)=1$ or $0$, then its state does not change over the
duration of the attractor; it is frozen. Alternatively, if
$0<\overline{O}(i)<1$, then node $i$ is active during the attractor.
\item Change the network topology by rewiring the connections to the
node chosen in the previous step. If it is frozen, then a new incoming
link from a randomly selected node $j$ is added to it. If it is
active, then one of its existing links is randomly selected and removed.
Note that this rewiring changes $K_i$.
\item The Boolean functions of network are regenerated.
Two different methods have been used:
\begin{itemize}
    \item Annealed model: A new Boolean function is generated for every
    node of the network.
    \item Quenched model: A new Boolean function is generated only for the chosen
   node $i$, while the others remain what they were previously.
\end{itemize}

\item Return to step 2.
\end{enumerate}

The time scale for an evolutionary change of the networks, steps
$2-6$ above, is called an epoch. As we will see, using this rewiring
rule the network topology evolves from a homogeneous one to a
heterogeneous one.
For simplicity, all random Boolean functions are generated with $p=1/2$, and 
therefore all Boolean functions with the same in-degree are equally likely to be generated. 

\begin{figure}[t]
\includegraphics[scale=0.35]{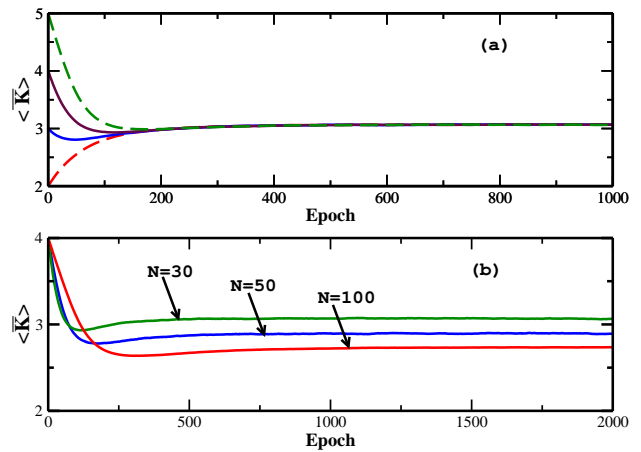}% Here is how to import EPS art
\caption{\label{AverageK}(color online) (a). Evolution of the ensemble averaged
in-degree connectivity in the annealed model for networks of size
$N=30$. The networks in each ensemble
initially start from different uniform connectivity,
$K_0 = 2, 3, 4$, and 5, but reach a same
statistical steady state $\langle \overline{K}\rangle=3.06$. Each
ensemble contained 15,000 realizations of the
network. (b). Evolution of ensemble averaged in-degree connectivity
for networks of three different size $N=30,50$, and 100 in the annealed
model.}
\end{figure}

\section{Simulation and Results }
We have simulated both the annealed and the quenched variants of the
model. Both variants give very similar results, and our principal
findings are the same for both variants. Therefore,  we will present
here mainly results from the annealed variant. Graph (a) of
Fig.~\ref{AverageK} shows the evolution of the average in-degree
connectivity $\overline{K}=\frac{1}{N}\sum_{i=1}^N K_i$ for networks
of size $N=30$ in the annealed variant of the model. Four curves are
shown. They show the results obtained by beginning with networks
with different uniform connectivity $K_0=2$, 3, 4, and 5. Each curve
is the average of 15,000 independent realizations of the network
evolution. This ensemble average is indicated by the angular
brackets. Each different realization in an ensemble begins with a
different random network and with a different random initial state
vector. Remarkably, despite the difference in initial conditions,
all four curves collapse after about 200 epochs, and they all
approach the same final statistical steady state that has an
average in-degree connectivity $\langle \overline{K}\rangle =
3.06$. This occurs without tuning and suggests that the final
evolved topology of the network is independent of the initial
topology of the network. The steady state value of $\langle \overline{K}
\rangle$ depends on the size of the system as shown in graph (b) of
Fig.~\ref{AverageK}. Starting with networks that all have the same
initial uniform connectivity $K_0=4$, but which have different size
$N=30,50$, and 100, we find that larger networks evolve to steady states
with smaller values of $\langle \overline{K} \rangle$. Very
similar results are obtained for the quenched version of the model.
For example, for networks with $N=30$ the steady state value of the
average connectivity is $\langle \overline{K}\rangle=3.08$.

\begin{figure}[t]
\includegraphics[scale=0.32]{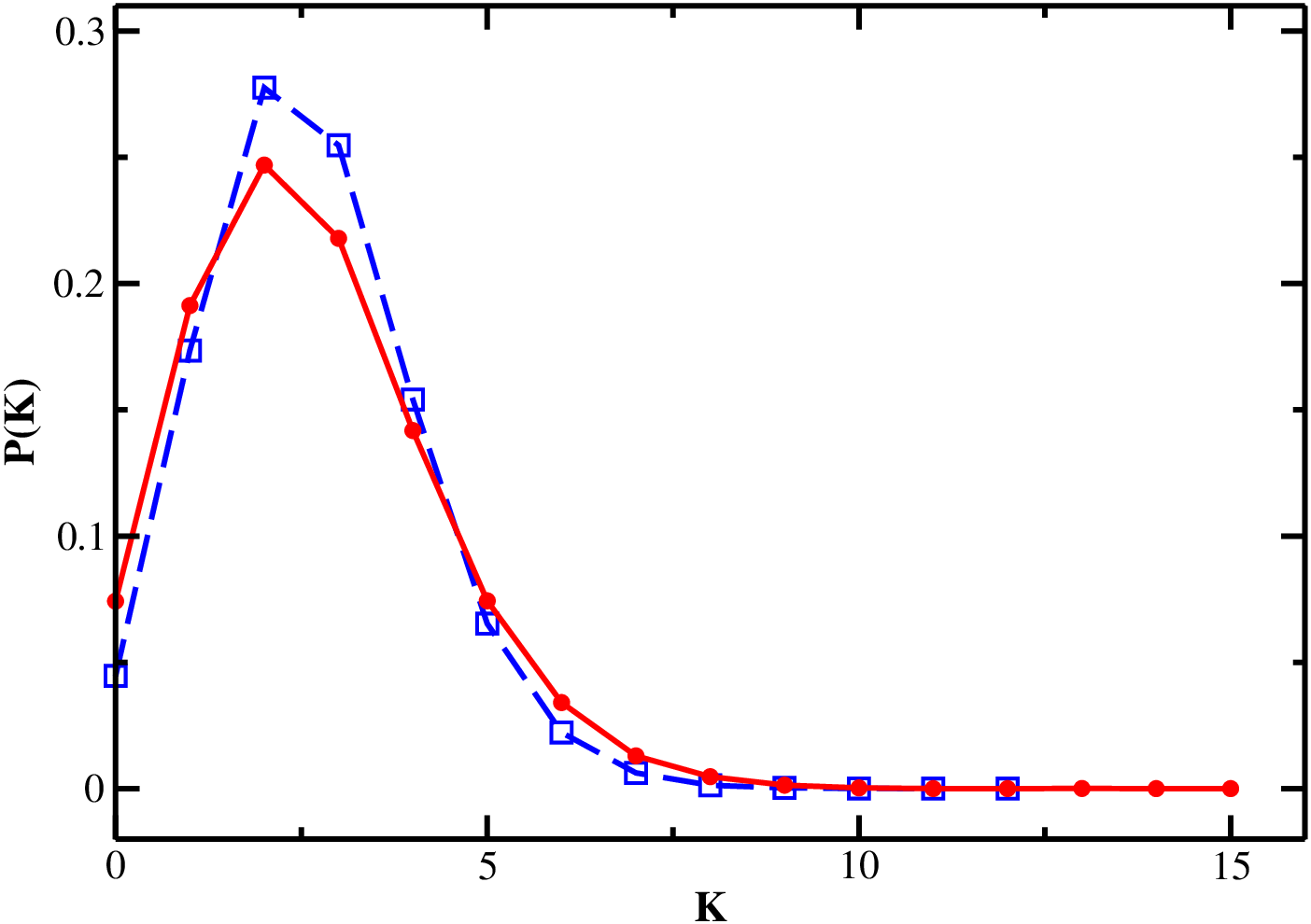}% Here is how to import EPS art
\caption{\label{degree200}(color online) Distribution of in-degree (square) and
out-degree (circle) connectivity in the annealed model. The size of
the networks is $N=200$.} \vspace{0.4in}
\includegraphics[scale=0.32]{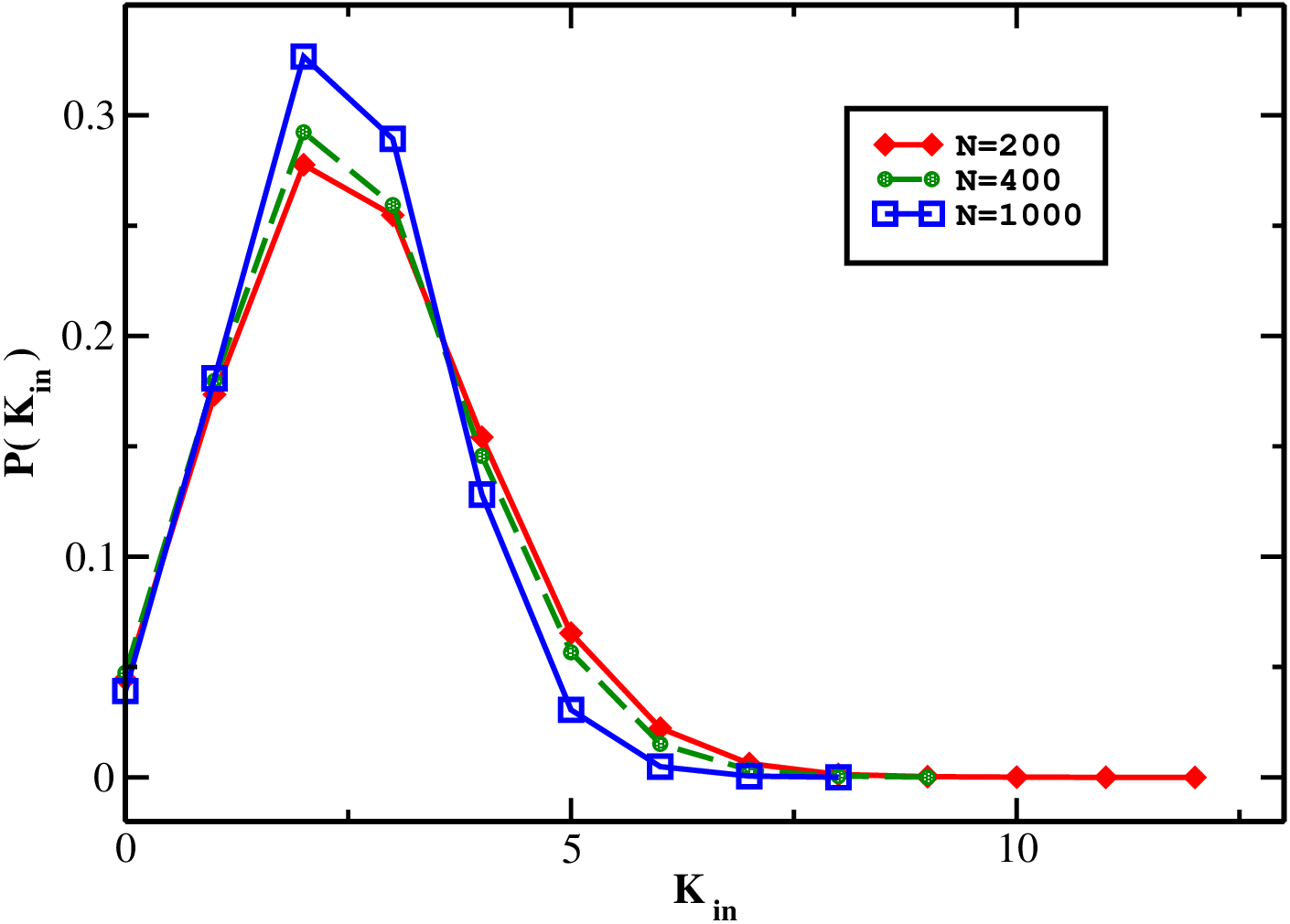}% Here is how to import EPS art
\caption{\label{degreesize}(color online) Distribution of in-degree connectivity
in the annealed model. The network sizes are $N=200$, $400$, and
$1000$.}
\end{figure}
We have also calculated the in-degree and out-degree connectivity
distributions, $P(K_{in})$ and $P(K_{out})$, of the evolved RBNs in
the steady state. Initially, all nodes of the network have a uniform
in-degree $K_0$, meaning that the in-degree distribution is a
discrete delta function $P(K_{in}) = \delta_{K_{in},K_0}$, and the
out-degree distribution is a binomial distribution. However, through
the evolutionary rewiring of the network both the in-degree and
out-degree distributions change. The in-degree and out-degree
distributions in the steady state of the annealed version, are shown
in Fig.~\ref{degree200} for $N=200$. They are both right skewed
bell-shaped distributions peaking at $K=2$. The out-degree
distribution remains a binomial distribution but the average
connectivity changes. The in-degree distribution, although it has
the same average connectivity as the out-degree distribution, is
more sharply peaked. As the size of the network grows, the in-degree
distribution becomes increasingly narrow and peaked at the value
$K_{in}=2$, as shown in Fig.~\ref{degreesize}. Based on this
observation, we conjecture that the distribution tends to converge
into a discrete delta function $\delta_{K_{in}, 2}$ in the large
network limit $N\rightarrow \infty$, indicating that the network
becomes a homogeneous RBN in that limit.

\begin{figure}[t]
\includegraphics[scale=0.32]{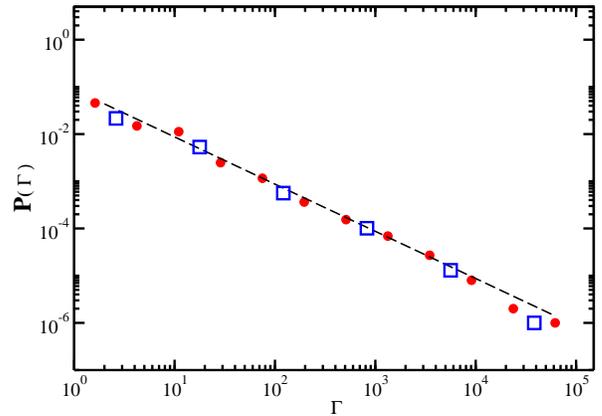}% Here is how to import EPS art
\caption{\label{powerlaw}(color online) Power law distribution of steady state
attractor period $\Gamma$ in both annealed (circle) and quenched
(square) models for $N=200$ system. The dashed straight line has a
slope of $1.0$. }
\end{figure}

In order to probe the dynamical nature of evolved steady states we
computed the distribution $P(\Gamma)$ of steady state attractor
period $\Gamma$ in the ensemble of RBNs simulated. The distribution
has a broad, power-law behavior for both the annealed and quenched
variants of the model. Figure~\ref{powerlaw} shows the results for
networks with $N=200$. As long as $N$ is about 30 or larger, results
for other size networks are similar. As discussed above, this
power-law distribution indicates that the networks have critical
dynamics. Also in the figure, the straight line has a slope of 1.0.
Thus, the critical exponent describing the power-law is
approximately 1.0. This value of the exponent is obtained for all
system sizes studied in both the quenched and annealed versions of
the model. In short, we find that a robust criticality emerges in
the evolutionary RBNs.

\begin{figure}[t]
\includegraphics[scale=0.32]{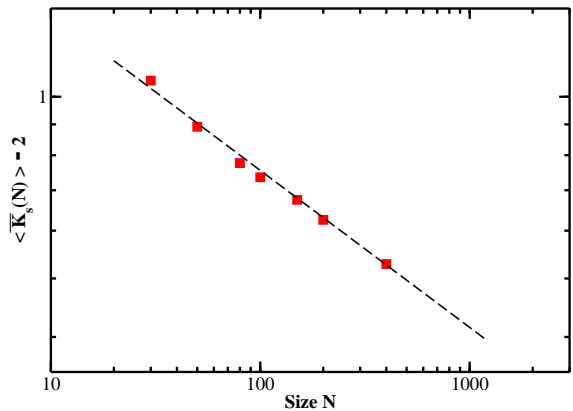}% Here is how to import EPS art
\caption{\label{finitesize}(color online) Finite-size effects in the annealed
model. The data shown are for systems of different sizes $N=30,
50, 80, 100, 150, 200$, and 400. The dashed straight line has a slop of
$0.26$.}
\end{figure}

Given the steady state value $\langle\overline{K} \rangle=2$ in
the large network limit $N\rightarrow\infty$, we studied the
finite-size effects in the model. As shown in Fig.~\ref{finitesize},
the values of $\langle \overline{K}(N)\rangle$ for finite $N$ obey
the scaling function
\begin{equation}\label{}
   \langle\overline{K}(N)\rangle - 2 = A N^{-\beta} .
\end{equation}
Fitting the data to this function, we find that the coefficient is
$A=2.50 \pm 0.06$ and the exponent is $\beta=0.264\pm 0.005$. Thus
the value of
$\langle\overline{K}(N)\rangle$ is always larger than 2 for finite
$N$. Note that steady state values of the average connectivity in
random threshold networks have a similar scaling form
\cite{Bornholdt}, but, in that case $A=12.4\pm 0.5$ and $\beta =
0.47 \pm 0.01$.

\section{Discussion and Conclusions}
The mechanism we find here that leads to the emergent critical state
has some similarity to self-organized criticality
(SOC)\cite{SOC1,SOC2}, but is different. SOC is the tendency of
driven dissipative dynamical systems to organize themselves into a
critical state far from equilibrium through avalanches of activity
of all sizes. In our particular model, the evolutionary Boolean
network is a dissipative dynamical system because multiple different
states may map into the same attractor so that information is lost.
Similar to SOC systems, our model is driven subject to two competing
rules, and the network organizes itself into a steady state that
results from a dynamical balance of the competition between those
rules. Moreover, the critical state is robust irrespective of
initial connectivity in both the quenched or annealed versions of
the model. The emergent critical state acts like a global attractor
in the evolution process. However, unlike the mechanism of
traditional SOC, but similar to the mechanisms that have been shown to
lead to criticality in random threshold networks \cite{Bornholdt} and
in homogeneous Boolean networks \cite{economics}, the
self-organizing mechanism here is based on a topological phase
transition in dynamical networks.

Our results indicate that, with the rewiring rules we use, networks
in the limit $N\rightarrow \infty$ will evolve to have a homogeneous
in-degree connectivity of 2. However, we find that finite size
networks evolve to have a broadly distributed heterogeneous
in-degree connectivity. The average in-degree connectivity of the
evolved networks is between 2 and 3 for biologically realized
network sizes. This result may be important for explaining the
observed structure of real gene regulatory networks. Real genetic
networks have a number of nodes $N$ ranging from near 100 to
thousands, and typically have a heterogeneous connectivity with an
average in-degree slightly larger than 2. For example, a recent
experiment studying the gene regulatory network of \emph{S.
cerevisiae} \cite{experiment2} found that the network contains 3420
genes and has an average in-degree connectivity $\overline{K}_{in} =
2.1$.

Finally, we note that real genetic networks exhibit an approximately
scale-free out-degree distribution while the in-degree distribution
is exponentially decaying\cite{experiment2,Albert}. In our numerical
study, we similarly obtain an in-degree distribution that decays
faster than the out-degree distribution, but our model does not
produce a scale-free like out-degree distribution. Therefore, in
order to be more realistic the model needs to be extended by adding
other factors that will capture this feature. We do not believe
though that such extensions will alter the principal conclusions of
this paper concerning the importance of finite size effects in the
evolution of network topology in real genetic networks.

\begin{acknowledgments}
We thank Maximino Aldana, Julio Monte, Toshimori Kitami  and Adam
Sheya for stimulating discussions. This work was supported
by the NSF through grant \#DMR-0427538, and by SI International
through the AFRL under contract \#FA8756-04-C-0258.
\end{acknowledgments}

\appendix*
\section{}
In the work presented here and in our previous studies\cite{economics,canalization,canalization2}
the following algorithm was used to determine the dynamical attractor.

Begin by using the initial state
$\mathbf{\Sigma}(0)$
as the ``checkpoint'' state.
Then for each time $T$, $0<T \leq T_1$, the state is updated using Eq.~1.
If 
$\mathbf{\Sigma}(T) = \mathbf{\Sigma}(0)$, 
then the attractor is found, it has period $T$,
and the search ends.
Note that the new state is compared only to the
checkpoint state and no other previous states.

If after $T_1$ updates no attractor is found then the checkpoint state
is changed to
$\mathbf{\Sigma}(T_1)$.
For each time $T$, $T_1<T\leq T_2$, the state is again updated using Eq.~1.
If 
$\mathbf{\Sigma}(T) = \mathbf{\Sigma}(T_1)$, 
then the attractor is found, it has period $T-T_1$,
and the search ends.

If after $T_2$ updates the attractor has still not been found then the checkpoint state
is changed to
$\mathbf{\Sigma}(T_2)$.
For each time $T$, $T_2<T\leq T_2 + T_{max}$, the state is again updated using Eq.~1.
If 
$\mathbf{\Sigma}(T) = \mathbf{\Sigma}(T_2)$, 
then the attractor is found, it has period $T-T_2$,
and the search ends.

Finally, if after $T_2+T_{max}$ updates the attractor has still not been found then the state
$\mathbf{\Sigma}(T_2+T_{max})$
is used as the new checkpoint state.
If 
$\mathbf{\Sigma}(T) = \mathbf{\Sigma}(T_2+T_{max})$, 
then the attractor is found, it has period $T-T_2-T_{max}$,
and the search ends.

If no attractor is found after this procedure, then we stop. In this case, the average output
state in Eq.~3 is calculated assuming that the attractor period is $\Gamma = T_{max}$ and that
the final checkpoint state
$\mathbf{\Sigma}(T_2+T_{max})$
is on the attractor.

This algorithm finds all attractors that have period less than or
equal to $T_{max}$ and that have a transient time to reach the
attractor from the initial state less than or equal to
$T_2+T_{max}$. For the results presented in this paper we used
$T_1=100$, $T_2=1000$, and $T_{max}=100 000$.

\bibliography{citation}% Produces the bibliography via BibTeX.

\end{document}